\documentclass[review]{elsarticle}

\usepackage{lineno,hyperref,amsfonts}
\modulolinenumbers[5]
\usepackage{amsmath}

\journal{Artificial Intelligence}

\begin{document}

\begin{frontmatter}

\title{Modeling Theory of Mind in Multi-Agent Games Using Adaptive Feedback Control}

\author[aff1,aff3]{Ismael T. Freire\corref{corresponding}}
\cortext[corresponding]{ITF and XDA equally contributed to this work.\\
Corresponding authors: ITF, PV}
\ead{ismaeltito.freire@gmail.com}

\author[aff1,aff2,aff3]{Xerxes D. Arsiwalla}

\author[aff1,aff3]{Jordi-Ysard Puigb\`o}

\author[aff1,aff2,aff3,aff4]{Paul Verschure\corref{corresponding}
\thanks{ITF and XDA equally contributed to this work.}}
\ead{pverschure@ibecbarcelona.eu}

\address[aff1]{Institute for Bioengineering of Catalonia (IBEC), Barcelona, Spain}
\address[aff2]{Universitat Pompeu Fabra (UPF), Barcelona, Spain}
\address[aff3]{Barcelona Institute of Science and Technology (BIST), Barcelona, Spain}
\address[aff4]{Catalan Institution for Research and Advanced Studies (ICREA), Barcelona, Spain}

\begin{abstract}
A major challenge in cognitive science and AI has been to understand how autonomous agents might acquire and predict behavioral and mental states of other agents in the course of complex social interactions. How does such an agent model the goals, beliefs, and actions of other agents it interacts with? What are the computational principles to model a Theory of Mind (ToM)? Deep learning approaches to address these questions fall short of a better understanding of the problem. In part, this is due to the black-box nature of deep networks, wherein computational mechanisms of ToM are not readily revealed. Here, we consider alternative hypotheses seeking to model how the brain might realize a ToM. In particular, we propose embodied and situated agent models based on distributed adaptive control theory to predict actions of other agents in five different game  theoretic tasks (Harmony Game, Hawk-Dove, Stag-Hunt, Prisoner’s Dilemma and Battle of the Exes). Our multi-layer control models implement top-down predictions from adaptive to reactive layers of control and bottom-up error feedback from reactive to adaptive layers. We test cooperative and competitive strategies among seven different agent models (cooperative, greedy, tit-for-tat, reinforcement-based, rational, predictive and other's-model agents). We show that, compared to pure reinforcement-based strategies, probabilistic learning agents modeled on rational, predictive and other's-model phenotypes perform better in game-theoretic metrics across tasks. Our autonomous multi-agent models capture systems-level processes underlying a ToM and highlight architectural principles of ToM from a control-theoretic perspective.  
\end{abstract}

\begin{keyword}
\texttt Multi-Agent Systems\sep Cognitive Architectures\sep Theory of Mind\sep Game Theory\sep Reinforcement Learning
\end{keyword}

\end{frontmatter}


\section{Introduction}
How do autonomous social agents model goals, beliefs and actions of other agents they interact with in complex social environments? This has long been a central question in cognitive science and philosophy of mind. It forms the basis of what is known as Theory of Mind (ToM), a long-standing problem in cognitive science concerned with how cognitive agents form predictions of  mental and behavioral states of other agents in the course of social interactions \cite{premack1978does,baron1985does}. The precise mechanisms by which the brain achieves this capability is not entirely understood. Furthermore, how a ToM can be embodied in artificial agents is very relevant for the future course of artificial intelligence (AI), especially with respect to human-machine interactions. What can be said concerning architectural and computational principles necessary for a Theory of Mind? As a small step in that direction, we propose and validate  control-based cognitive architectures to predict actions and models of other agents in five different game theoretic tasks. 

It is known that humans also use their ToM to attribute mental states, beliefs, and intentions to inanimate objects \cite{premack1990infant}, suggesting that the same mechanisms governing ToM may be integral to other aspects of human cognition. We argue that in order to understand both, how ToM is implemented in biological brains and how it may be modeled in artificial agents, one must first identify its  architectural principles and constraints from biology. 

Several notable approaches addressing various aspects of this problem already exist, particularly, those based on artificial neural networks \cite{lanctot2017unified,lerer2018learning}. Recent work in  \cite{rabinowitz2018machine} on Machine Theory of Mind is one such example. This work proposes that some degree of theory of mind can be \emph{autonomously} obtained from pure Reinforcement Learning (RL). However, the use of Black-Box Optimization algorithms, hinders the understanding one can obtain at the mechanistic level. Because BBO algorithms are able to approximate any complex function (as suggested for Long-Short Term Memory (LSTM)-based neural networks \cite{schmidhuber2015deep}), one cannot use that to decipher specific mechanisms that may underlie a ToM in these systems. 

Another interesting approach to ToM follows the work of \cite{yoshida2008game,baker2011bayesian,baker2017rational,tenenbaum2018building}  which uses hierarchical Bayesian inference. These methods are cognitively-inspired and suggest the existence of a "psychology engine" in cognitive agents to process ToM computations. Nevertheless, the challenge for this approach remains explaining how the computational cost of running these models might be implemented in biological substrates of embodied and situated agents. 

A third approach to this problem follows from work on autonomous multi-agent models (see  \cite{albrecht2018autonomous} for a recent survey). This approach has had its roots in statistical machine learning theory and robotics. It includes agent models capable of policy reconstruction, type-classification, planning, recursive reasoning and group modeling. In a sense, this is closest to the approach we take in this work, even though, our models are grounded in cognitive control systems. Open challenges in the field of autonomous multi-agent systems include modeling fully embodied agents that operate with only partial observability of their environment and are flexibly able to learn across tasks (including meta-learning), when interacting with multiple types of other agents.  

The cognitive agent models we propose here, advance earlier work on Control-based Reinforcement Learning (CRL) \cite{freire2018modeling}, where we studied the formation of social conventions in the Battle of the Exes game. The CRL model implements a feedback control loop handling the agent's reactive behaviors (pre-wired reflexes), along with an adaptive layer that uses  reinforcement learning to maximize long-term reward. We showed that this model was able to simulate human data in the above-mentioned coordination game. 

The new contribution of the current paper is to advance real-time learning and control strategies to model agents with specific phenotypes, that can learn to either model or predict the opposing agents actions.  Using top-down predictions from the adaptive to reactive control layers and bottom-up error feedback from reactive to adaptive layers,  we test cooperative and competitive strategies for seven different multi-agent behavioral models. The purpose of this study is to understand how the architectural assumptions behind each of these models impact agent performance across standard game-theoretic tasks, and what this implies for the development of an artificial embodied ToM.

\section{Methods}

\subsection{Game Theoretic Tasks}

Benchmarks inspired by game theory are becoming standard in the multi-agent reinforcement learning literature \cite{lanctot2017unified,lerer2018learning,rabinowitz2018machine,freire2018limits}. However, most of the work developed in this direction presents models that are tested in one single task or environment \cite{kleiman2016coordinate, perolat2017multi, peysakhovich2018prosocial, freire2018modeling, freire2018modelingb,gasparrini2018loss}, at best two \cite{leibo2017multi, peysakhovich2017consequentialist}. This raises a fundamental question about how these models generalize to deal with a more general or diverse set of problems. Therefore, this approach does not readily enable one to extract principles and mechanisms or unravel the dynamics underlying human cooperation and social decision-making.


In this work, we want to go a step further and propose a five-task benchmark for predictive models based on classic normal-form games extracted from Game Theory: the Prisoner's dilemma, the Harmony game, the Hawk-dove, the Stag-hunt and the Battle of the Exes.

In its normal-form, games are depicted in a matrix that contains all the possible combinations of actions that players can choose, along with its respective rewards.
The most common among them are the dyadic games known as social dilemmas, in which players have to choose between two actions: cooperate or defect. One action (cooperate) is more generous and renders a good amount of reward to each player if both choose it, but gives very poor results if the other player decides to defect. On the other hand, the second action (defect) provides a significant individual reward if its taken alone, but a very small one if both choose it. 

Although simple in nature, these dyadic games are able to model key elements of social interaction, such as the tension between the benefit of a cooperative action and the risk (or temptation) of free-riding. This feature can be described in a general form, such as:

\begin{table}[h]
\begin{center}
\begin{minipage}[h]{0.5\linewidth} 
\centering
\begin{tabular}{|c|c|c|}
\hline
 & Cooperate & Defect   \\  \hline
Cooperate & R, R  &  S, T   \\  \hline
Defect & T, S &  P, P   \\ \hline
\end{tabular}
\caption{ A Social Dilemma in matrix-form }
\end{minipage}
\end{center}
\end{table}

This matrix represents the outcomes of all possible combination of actions between the row player and the column player. R stands for "reward for mutual cooperation"; T for "temptation of defecting"; S for the "sucker's payoff for non-reciprocated cooperation"; and P for "punishment for mutual defection". By manipulating the relationship between the values of this matrix, many different situations can be obtained that vary in terms of what would be an optimal solution (or Nash equilibria \cite{nash1950equilibrium}).

The Prisoner's Dilemma (table 2) represents the grim situation in which the temptation of defecting (T) is more rewarding that mutual cooperation (R), and the punishment for mutual defection (P) is still more beneficial than a failed attempt of cooperation (S). This relationship among the possible outcomes can be stated as $T > R > P > S$. Mutual defection is the only pure Nash equilibrium in this game since there is no possibility for any player to be better off by individually changing its own strategy.

On the Stag-Hunt (table 3) we face a context in which mutual cooperation (R) gives better results and individual defection (T), but at the same time a failed cooperation (S) is worse than the punishment for mutual defection (P). In this case, $R > T > P > S$, there are two Nash equilibria: mutual cooperation and mutual defection.

Hawk-dove (table 4) presents a scenario in which temptation (T) is more rewarding than cooperation (R), but a mutual defection (P) is less desirable than non-reciprocated cooperation (S). So, for a relationship of $ T > R > S > P$ like this one, there are three Nash equilibria: two pure anti-coordination equilibrium, in which each player chooses always the opposite action of its opponent, and one mixed equilibrium, in which each player probabilistically chooses between the two pure strategies.

The fourth type of social dilemma is the Harmony game (table 5). In this case ($R > T > S > P$), the game has only one pure equilibrium, pure cooperation, since mutual cooperation (R) renders a better outcomes that the temptation to defect (T), and also the penalty for failing to cooperate (S) is less than the punishment for mutual defection (P).

Finally, as the last game, we will introduce a coordination game, the Battle of the Exes \cite{hawkins2016formation,hawkins2018emergence}. In this type of game, the main goal is to achieve coordination between two players (either congruent coordination on the same action or incongruent coordination upon choosing different actions), since the failure to do so is heavily penalized (see table 6). Following the previous nomenclature applied in the social dilemmas, we could define this game as $T > S$; $R, P = 0$. This game has two pure-dominance equilibria, in which one player chooses the more rewarding action and the other the low rewarding action; and a turn-taking equilibrium, in which players alternate over time in choosing the more rewarding action.

\begin{table}[h]
\begin{center}
\begin{minipage}[h]{0.5\linewidth} 
\centering
\begin{tabular}{|c|c|c|}
\hline
 & Cooperate & Defect   \\  \hline
Cooperate & 2, 2  &  0, 3   \\  \hline
Defect & 3, 0 &  1, 1   \\ \hline
\end{tabular}
\caption{ prisoner's dilemma }
\vspace{.5cm}
\begin{tabular}{|c|c|c|}
\hline
 & Cooperate & Defect   \\  \hline
Cooperate & 3, 3  &  0, 2   \\  \hline
Defect & 2, 0 &  1, 1   \\ \hline
\end{tabular}
\caption{ stag-hunt } 
\end{minipage}%
\begin{minipage}[h]{0.5\linewidth} 
\centering
\begin{tabular}{|c|c|c|}
\hline
 & Cooperate & Defect   \\  \hline
Cooperate & 2, 2  &  1, 3   \\  \hline
Defect & 3, 1 &  0, 0   \\ \hline
\end{tabular}
\caption{ hawk-dove } 
\vspace{.5cm}
\begin{tabular}{|c|c|c|}
\hline
 & Cooperate & Defect   \\  \hline
Cooperate & 3, 3  &  1, 2   \\  \hline
Defect & 2, 1 &  0, 0   \\ \hline
\end{tabular}
\caption{ harmony game } 
\end{minipage}
\end{center}
\end{table}

\begin{table}[h]
\begin{center}
\begin{minipage}[h]{0.5\linewidth} 
\centering
\begin{tabular}{|c|c|c|}
\hline
 & A & B   \\  \hline
A & 0, 0  &  1, 4   \\  \hline
B & 4, 1 &  0, 0   \\ \hline
\end{tabular}
\caption{ battle of the exes }
\end{minipage}
\end{center}
\end{table}


This selection of games provides enough variability to test how learning agents can perform across different contexts, so we avoid problems derived from over-fitting on a specific payoff distribution, or related to the possibility of a model to exploit/capitalize on certain features of a game that could not have been predicted beforehand. Moreover, a similar selection of games have been tested in human experiments, proving to be sufficient to classify a reduced set of behavioral phenotypes across games in human players \cite{poncela2016humans,sanfey2007social}. Using the above games as benchmarks, below we describe the control-theoretic framework and seven agent models that capture phenotypes relevant for testing ToM capabilities across games.

\subsection{Control-Based Reinforcement Learning}
\begin{figure*}[h]
\centering
\includegraphics[width=.5\textwidth]{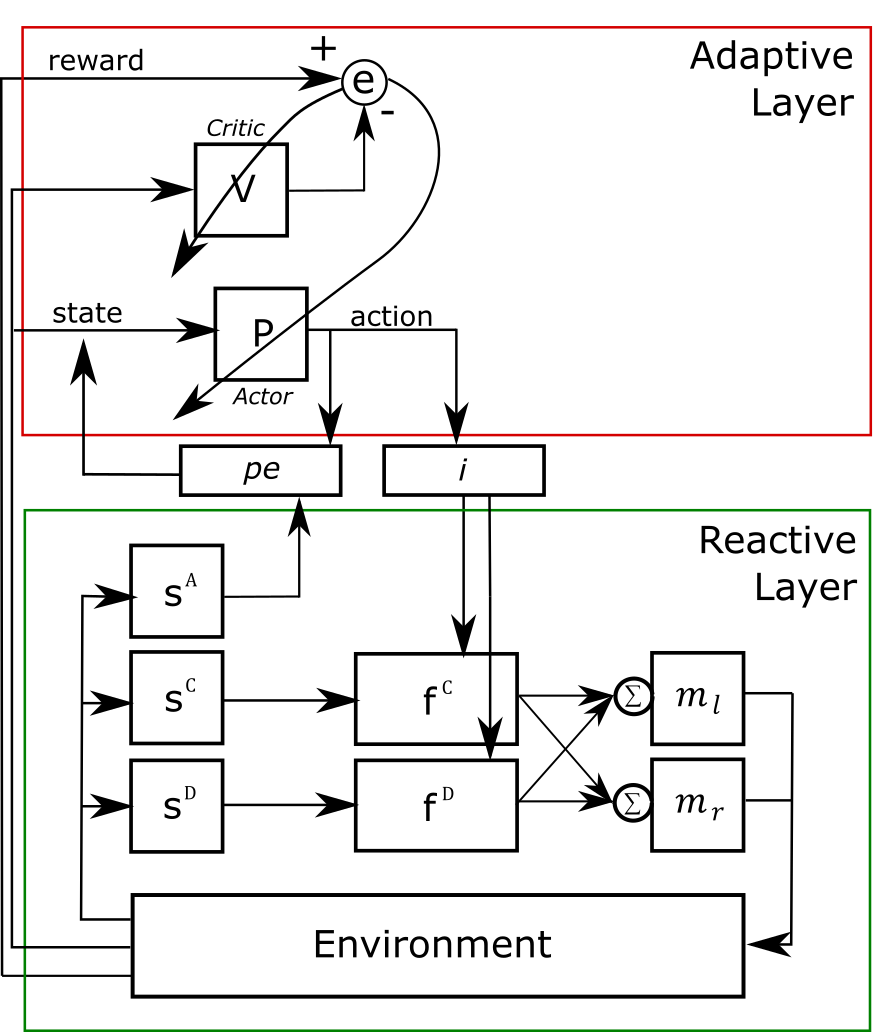}
\caption{A) Representation of the Control-based Reinforcement Learning (CRL) model. The top red box depicts the Adaptive Layer or the original CRL model, composed of an Actor-Critic Temporal-Difference (TD) learning algorithm. The bottom green box represents the Reactive layer, with its three sets of sensors, one for the 'cooperate' location (sC), one for the 'defect' location (sD), and one for the other agent (sA); the two reactive behaviors, \textit{"approach cooperate location"} (fC), \textit{"approach defect location"} (fD); and the two motors, one for the left wheel (ml) and one for the right wheel (mr). Between the two layers, the inhibitory function (i) regulates which reactive behaviors will be active depending on the action received from the Adaptive layer, while the error monitoring function (pe) manages the mismatch between the opponent's predicted behavior and the actual observation in real-time.}
		\label{CRL}
\end{figure*}

Our starting point is the Control-based Reinforcement Learning (CRL) model presented in previous work \cite{freire2018modeling}. The CRL is a biologically grounded cognitive model composed of two control layers (\textit{Reactive} and \textit{Adaptive}, see Figure \ref{CRL}), and based on the principles of the Distributed Adaptive Control (DAC) theory \cite{verschure2003environmentally}. The \textit{Reactive Layer} represents the agent's pre-wired reflexes and serves for real-time control of sensorimotor contingencies. The \textit{Adaptive Layer} endows the agent with learning abilities that maximize long-term reward by choosing which action to perform in each round of the game. This layered structure allows for top-down and bottom-up interactions between the two layers \cite{moulin2016top}, resulting in an optimal control at different time-scales: within each round of play and across rounds \cite{freire2018modeling}.

The \textit{Reactive Layer} is inspired by Valentino Braitenberg's Vehicles \cite{braitenberg1986vehicles} and presents the intrinsic mechanic behaviors that are caused by a direct mapping between the sensors and the motors of the agent. The main one, \textit{"orienting towards rewards"}, is based on a cross excitatory connection and a direct inhibitory connection between the reward sensors and the two motors. This results in an approaching behavior in which the agents will turn towards a reward location increasingly faster the closer they detect it. Since the agents are equipped with two sets of sensors specifically tuned for detecting each reward location, they also have two instances of this reactive behavior, one for the 'cooperative' and one for the 'individual' reward (see Figure 1 green box). 

The \textit{Adaptive Layer} is in charge of prediction, learning, and decision-making. Each of the Agent Models described in the next section instantiates a different Adaptive layer. Its main role is to learn from previous experience and to decide at the beginning of each round which action the agent will take. When playing normal-form games, the Adaptive Layer receives as inputs the state of the environment, which corresponds with the outcome of the last round of the game, and the reward obtained because of that final state. This information is used to produce an action as the output. The states $S$ can be either 'R', 'S', 'T' or 'P' (see Table 1), and determines what is the reward that the agent will obtain based on the specific payoff matrix of the game being played. The actions $A$ are two: \textit{"cooperate"} or \textit{"defect"}.

Depending on the action that the \textit{Adaptive Layer} selects, there is a top-down selective inhibition that affects the reactive behaviors of the \textit{Reactive Layer}. If the action selected is \textit{"cooperate"}, the \textit{"approach individual reward"} reactive behavior is inhibited, thus focusing the agent's attention only on the cooperative reward and on the other agent. Conversely, if the action selected is \textit{"defect"}, the reactive behavior inhibited will be \textit{"approach cooperative reward"}. This mechanism aims to mimic how biological systems execute top-down control over a hierarchy of different control structures \cite{corbetta2002control, koechlin2003architecture, munakata2011unified}.

The error monitoring function works in real-time as a bottom-up mechanism that signals an error function from the Reactive layer sensors to the Adaptive layer decision-making module. The error signal is only triggered when the agent detects an inconsistency between its initial prediction of the opponent's action and the real-time data obtained by its sensors. If a prediction error occurs, the error monitoring function will update the current state of the agent and this will make the decision-making module output a new action. Along with the top-down inhibitory control, this module is inspired by evidence from cognitive science about the role of bottom-up sensory stimuli in generating prediction errors \cite{corbetta2002control,den2012prediction, wacongne2011evidence}.

\subsection{Agent Models}
\subsubsection{Original Model}
\begin{figure}[ht]
\centering
\includegraphics[width=.95\textwidth]{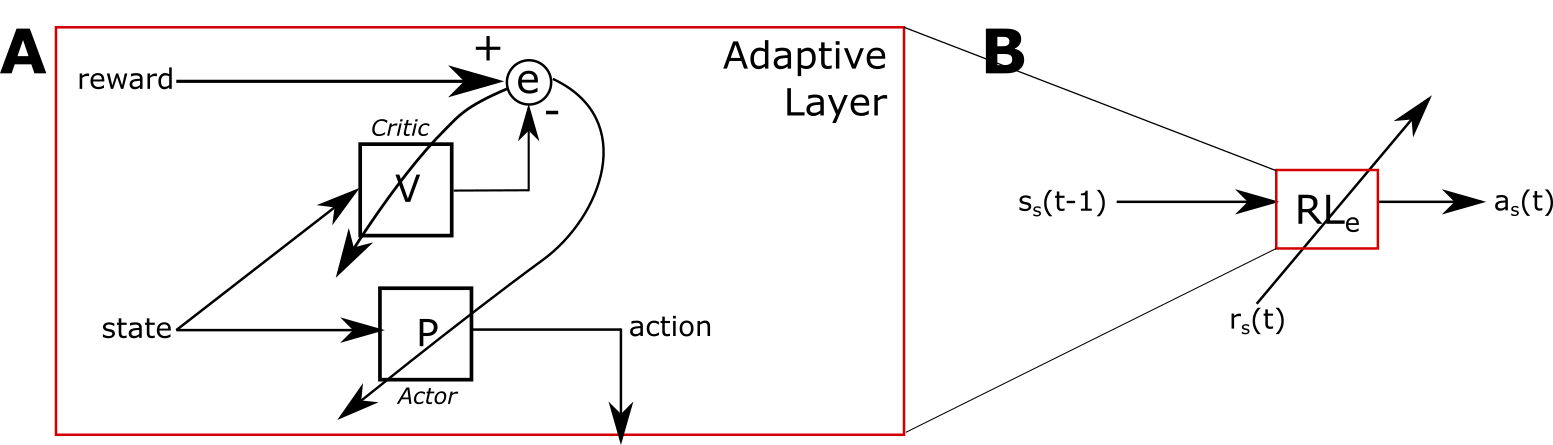}
\caption{Representations of the Original model. Panel A shows a detailed representation of the algorithm components, as implemented in the Adaptive layer of the original implementation of the CRL model \cite{freire2018modeling}. Panel B shows a compressed representation showing only the inputs and outputs of that same model.}
\label{OriginalModel} 
\end{figure}

The Original CRL model uses an Actor-Critic version of the Temporal-Difference (TD) learning algorithm \cite{sutton1988learning} for maximizing long-term reward (see \cite{freire2018modeling} for a detailed description of the implementation). 

In brief, the TD-learning algorithm selects an action according to a given policy $P (a=a_t|s=s_{t-1})$ . Once a round of play is finished, the reward $r(s_t)$ obtained by the agent will update the TD-error $e$ signal, following: $e (s_{t}) = r (s_t) + \gamma  V_{\Pi} (s_t) - V_{\Pi} (s_{t-1})$ where $\gamma$ is a discount factor 
and $V_{\pi} (s_{t})=\gamma r(s_t)$ is the Critic. Finally, the policy (or Actor) will be updated following $\Pi (a_t, s_{t-1}) = \Pi (a_t, s_{t-1}) + \delta e (s_{t-1})  $, where $\delta$ is a learning rate that is set to $0.15$. 


\subsubsection{Rational Model}
\begin{figure}[ht]
\centering
\includegraphics[width=.5\textwidth]{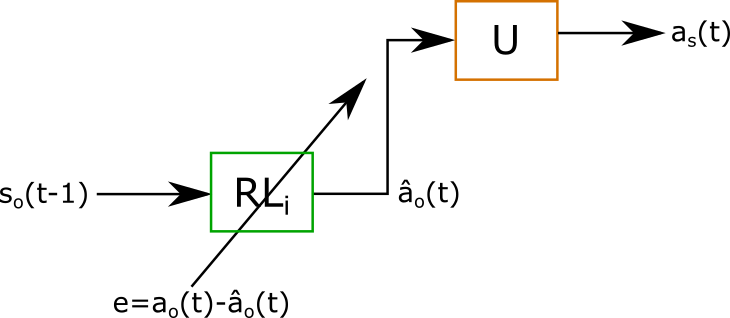}
\caption{Representation of the Rational Model. This model is composed of a predictive module (RL) that learns to predict the opponent's future action and a utility maximization function (U) that computes the action that yields the highest reward based on the opponent's predicted action. At the end of each round, the RL module is updated based on its prediction error.}
\label{RationalModel}
\end{figure}

The Rational Model is a predictive model that represents an ideal perfectly-rational and self-interested player provided that its predictions are correct. Its function is to serve as a benchmark for the other predictive models since once it learns to predict its opponent's actions accurately, it will always respond automatically with the best response to that predicted action. It is composed of two main functions: a prediction module and a deterministic utility maximization function (see Figure \ref{RationalModel}). The first module tries to predict the next action of the opponent by using a TD-learning algorithm that uses the opponent's previous state as an input. Once the prediction is made, the second module calculates the action that will render the highest reward assuming that the opponent has chosen the predicted action. 

This predictive model uses the same TD-learning algorithm as the original, but substitutes the explicit reward for an implicit reward that comes from the error in predicting the other agent's action. We call this reward 'implicit' because is calculated internally and it is based on whether the action of the opponent was accurately predicted or not. That way, an accurate prediction will render a positive reward, while an incorrect one will receive a negative one. 

The implicit reward signal is calculated by the function:

\begin{equation}
  R_i= 
\begin{cases}
  1,    & \text{if } A_o{t} = \hat{A}_o{t-1} \\
  -1,   & \text{otherwise}
\end{cases}
\end{equation}

\subsubsection{Predictive Model}
\begin{figure*}[h]
\centering
\includegraphics[width=.5\textwidth]{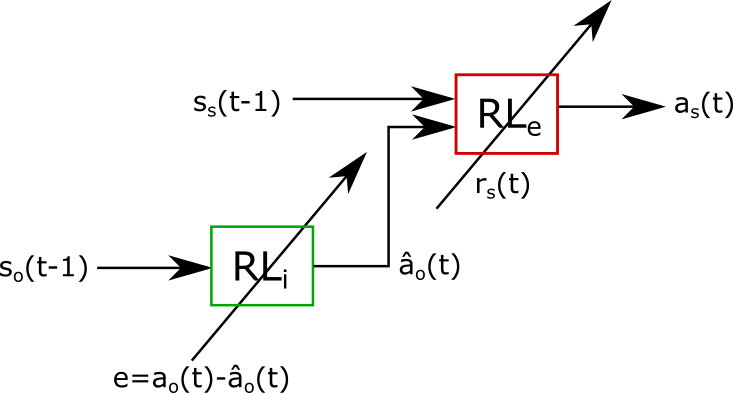}
\caption{Representation of the Predictive Model. This model is composed of a predictive module (RL) that learns to predict the opponent's future action and a TD learning module (RL) that uses that prediction along with the previous state to learn the optimal policy. At the end of the round, the predictive-RL (green) is updated according to its error in the prediction.}
\label{PredictiveModel}
\end{figure*}
The \textit{Predictive} model has two distinct RL modules: one that learns to predict and one that chooses an action based on that prediction  (see Figure \ref{PredictiveModel}). The predictive RL algorithm is same as the one used by the Rational agent described above, so it also learns from the implicit reward obtained by Equation 1. The prediction generated at the beginning of each round by the predictive-RL is sent as an input to the second RL algorithm. This one, in turn, uses the combined information of the opponent's predicted action and the state of the previous round to learn the optimal action. For its update function, the second RL algorithm uses the explicit reward obtained in that round of the game.

\subsubsection{Other's Model}
\begin{figure}[h]
\centering
\includegraphics[width=.5\textwidth]{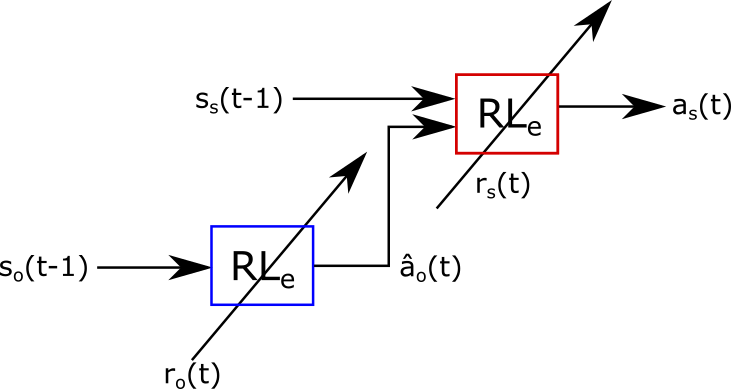}
\caption{Representation of the Other's-Model. This model is composed by two TD learning algorithms. The first one (left, blue) learns to predict the opponent's future action while the second (right, red) uses a that prediction to learn the optimal policy. The first algorithm is updated with the opponent's reward and the second with the agent's own reward.}
\label{OtherModel}
\end{figure}
The \textit{Other's Model} is also composed by RL algorithms, one predictive and one for learning the optimal policy (see Figure \ref{OtherModel}. Technically, the second RL module is identical to the one used by the Predictive model: it integrates both the previous state of the game and the opponent's predicted action in order to learn the optimal policy. However, it differs from the Predictive model in the way its first module is designed. In this case, the predictive-RL algorithm is updated by the explicit reward that the opponent has obtained, in an attempt to create an internal model of the other agent's policy. At the functional level is very similar to the Predictive model, but with an important difference: while the Predictive model tries predict the opponent's action by focusing on its overt behavior, the Other's Model does it trying to learn the internal policy of its opponent.

\subsubsection{Deterministic Agent Models }
In order to test the correct functioning of the predictive capacities of the ToM agents described above, we have developed a number of agents have a fixed behavior or policy. The first two represent two behavioral phenotypes observed in humans \cite{poncela2016humans} while the third one is a classic benchmark of Game Theory \cite{axelrod1981evolution}.

\paragraph{Greedy Agent Model}
This model implements a simple behavioral strategy of pure self-utility maximization: it always chooses the action the renders the highest reward to itself, without taking into account the opponent's action. So for the games described in this work, it will always choose to 'defect' with the exception of the Stag-Hunt and the Harmony Game, where the cooperative reward can give a higher reward than the temptation to defect ($R > T$).

\begin{equation}
  \pi_g= 
\begin{cases}
  cooperate,    & \text{if } R > T \\
  defect,   & \text{otherwise}
\end{cases}
\end{equation}

\paragraph{Cooperative / Nice Agent Model}
As a deterministic counterpart of the Greedy model, the Nice model executes an even simpler deterministic strategy: it will always choose cooperation. 
\begin{equation}
\pi_n = P (a=\text{cooperate}|s_t) = 1
\end{equation}

\paragraph{Tit-for-Tat Agent Model}
This simple yet powerful strategy became popular in the famous Axelrod's tournament \cite{axelrod1980effective} where it won to all competing algorithms and strategies in a contest based on the Iterated Prisoner's Dilemma. We have introduced this opponent because we expect that this agent will give problems to the predictive models. Tit-for-tat starts always by cooperating, and from then on it always chooses the last action made by its opponent. Since its capacity to switch actions does not depend on a specific policy, it cannot be learned by just taking into account variables such as its previous state or its reward. It can be described by the following equation, where $a_\text{opponent}$ is the action made by the opponent at $t-1$:
\begin{equation}
  \pi_g= 
\begin{cases}
  cooperate,    & \text{if } t = 0 \\
  a_\text{opponent} (t-1),   & \text{otherwise}
\end{cases}
\end{equation}

\subsection{Experimental Setup}

First, we perform four different experiments in which we test the four learning models described above (Original, Rational, Predictive, Other's-Model) in each of the five games described in section 2.1 (Prisoner's Dilemma, Hawk-Dove, Stag-Hunt, Harmony game and Battle of the Exes) and against opponents of different level of complexity (Greedy, Nice, Tit-for-tat and Original), resulting in a 4x5x4 experimental setup.

This way, in experiment one, all models are tested against the Greedy agent; in experiment two, against the Nice agent; in experiment three, against the Tit-for-tat agent; and in experiment four, against the Original agent. 

In all experiments, each model plays 50 times each of the five games, during a total amount of 1000 rounds per iteration (i.e., 5 games, 50 dyads per game, 1000 rounds per dyad). On each iteration, every model starts learning from scratch, with no previous training period.

After that, we perform a fifth experiment to study in a real-time spatial version of the games, special cases in which the models encountered problems in the previous setting. So, in this first four experiments, we use the discrete-time version of the games, and for the fifth experiment, a real/continuous time version of the tasks. 

In the discrete-time version, both agents simultaneously choose an action at the beginning of each round. Immediately after that, the outcome of the round is calculated and each agent receives a reward equal to the value stated in the payoff matrix of the game. 

\begin{figure}[h]
\centering
\includegraphics[width=\textwidth]{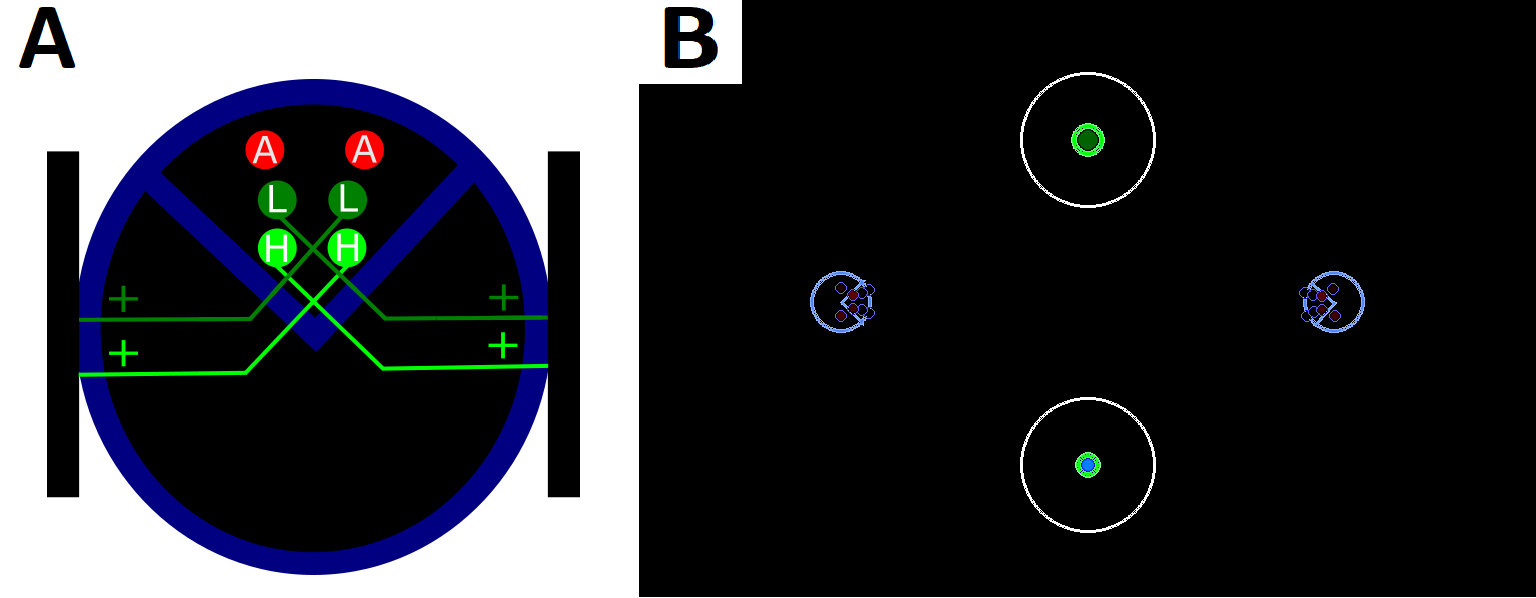}
\caption{A) A top-down visualization of the agents used in the continuous version of the games. The green circles represent the location-specific sensors, and the red circles the agent-specific ones. The green lines that connect the location sensors with the wheels represent the Braitenberg-like excitatory and inhibitory connections. \space 
B) Image of initial conditions in the continuous/real-time spatial version of the tasks. The blue circles are the two agents facing each other (representing two ePuck robots viewed from the top). The big green circle represents the 'cooperate' reward location; the small green circle, the 'defect' reward. The white circles around each reward spot mark the threshold of the detection area.}
\label{Expsetup}
\end{figure}

In the continuous-time version, the two agents can choose to go to one of the two equally distant locations that represent the two actions of a 2-action matrix-form game (see Figure \ref{Expsetup}-B). In this version, agents also choose an action at the beginning of the round, but then they have to navigate towards the selected location until one of them reaches its destination. The real-time dimension of this implementation allows the possibility of changing the action during the course of the round as well as to receive feedback in real-time. Each round of the game begins with both agents in their initial positions, as shown in Figure \ref{Expsetup}-B. When the agents reach a reward location, the round ends, the rewards are distributed accordingly to the payoff matrix and the game restarts with the agents back in their starting positions. 


The agents of the continuous version are embodied in virtual ePuck robots (see \ref{Expsetup}-A. They are equipped with two motors (for the left and right wheels) and three pairs of proximity sensors; one pair is specialized in detecting the other agent, and the other two in detecting the two distinct reward locations ('cooperate' or 'defect').


\section{Results}
In order to study the performance of the four agents (Original, Rational, Predictive and Other's-Model), we focus on three aspects of the interaction: \textit{Efficacy}, \textit{Prediction Accuracy}, and \textit{Stability}. 

\textit{Efficacy} tells us how competent the models were in obtaining rewards on average and in relation to each other. It's computed by calculating the mean score per game for each model. \textit{Stability} measures how predictable the behavior was or, equivalently, whether the models converged to a common strategy or alternated between non-deterministic states. In other words, stability quantifies how predictable are the outcomes of the following rounds based on previous results by using the information-theoretic measure of surprisal (also known as self-information), which can be defined as the negative logarithm of the probability of an event \cite{shannon1948mathematical}. By computing the average surprisal value of a model over time, we can visually observe how much did it take for a model to converge to a stable strategy and for how long it was able to maintain it. Finally, \textit{Prediction Accuracy} give us a better understanding of how well the predictive models were able to actually predict their opponent's behavior. It's measured by the average accumulated prediction error of each predictive model in each game. 

\subsection{Experiment 1: Against a deterministic-greedy agent}
\begin{figure*}[ht]
\centering
\includegraphics[width=.90\textwidth]{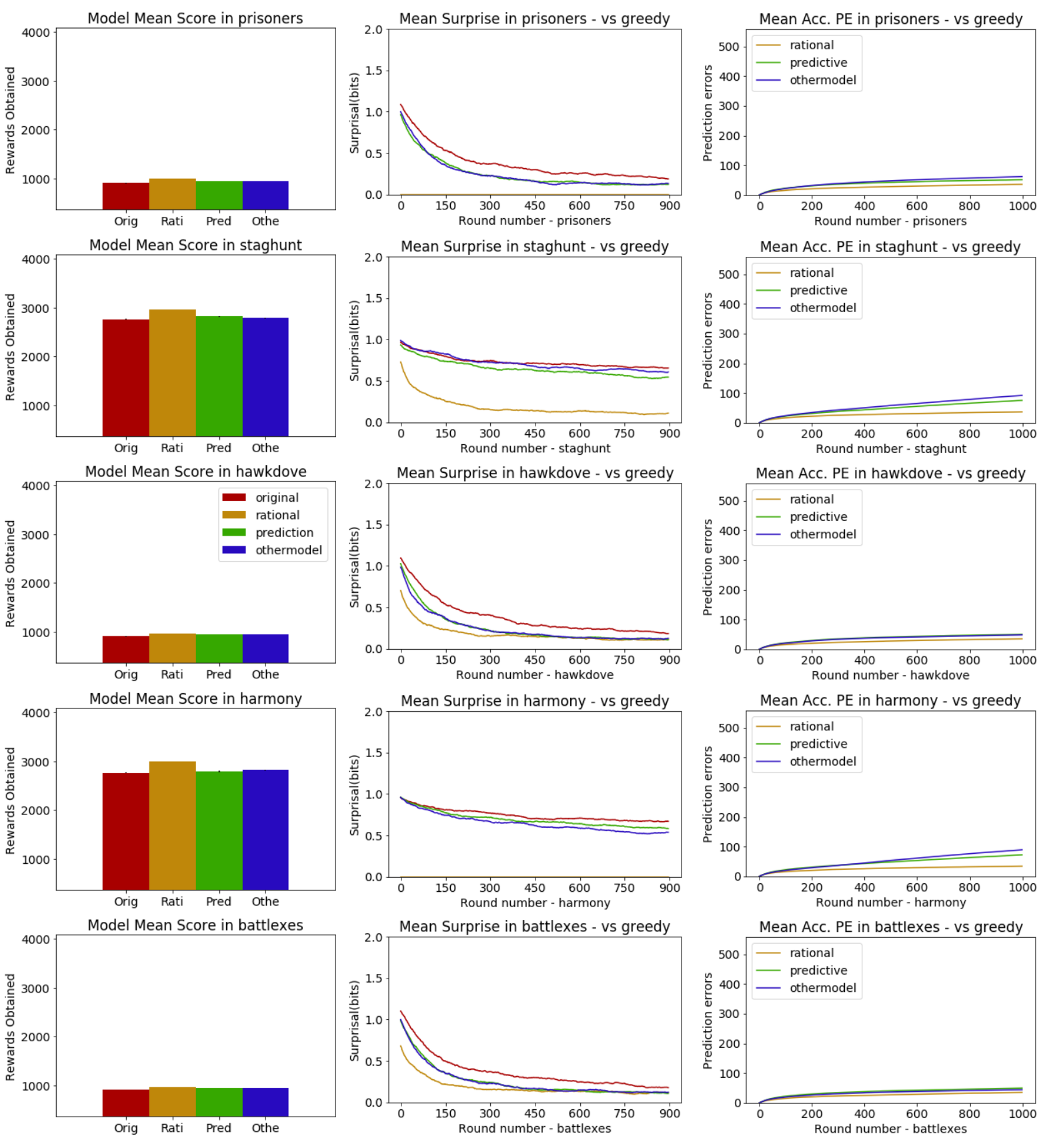}
\caption{Main results of the four behavioral models (Original, Rational, Predictive and Other's) against a Greedy agent.}
\label{results_greedy}
\end{figure*}

The results of this experiment show that on average all models were equally effective since they obtained a similar amount of rewards (see Figure \ref{results_greedy}). The Rational model only achieves a slightly better performance than the rest in the Stag-hunt and the Harmony game, so this is a sign that the rest of the models were performing almost optimally. Of course, if we pay attention to the overall score obtained among the five games, we can observe a significant difference between the performance obtained by all models in the Harmony game and the Hawk-dove when compared to the other three games. This salient difference is not due to a malfunctioning of the predictive models, as we can see by the results in Stability and Prediction Accuracy. It is caused by the constraints imposed by the Greedy opponent strategy in those games were $T > R$, since are the ones in which it always select to 'defect'. We see that all models learn to predict or to adapt their strategy to this opponent. In terms of surprisal, we can observe how Rational Agents outperforms all the other models. This result is not \textit{surprising} given the simplicity of the opponent's strategy. Against these simple predictable agents,  we consider the Rational agent a reference point of optimal performance, against which the other models can be compared. In this line, a more significant result is that overall, both Predictive and Other's-Model converge to a stable state faster than the original model, who lacks the ability to predict the opponent's action. In terms of Prediction Accuracy, no significant difference is observed. 

\subsection{Experiment 2: Against a deterministic-nice agent}
\begin{figure*}[ht]
\centering
\includegraphics[width=.90\textwidth]{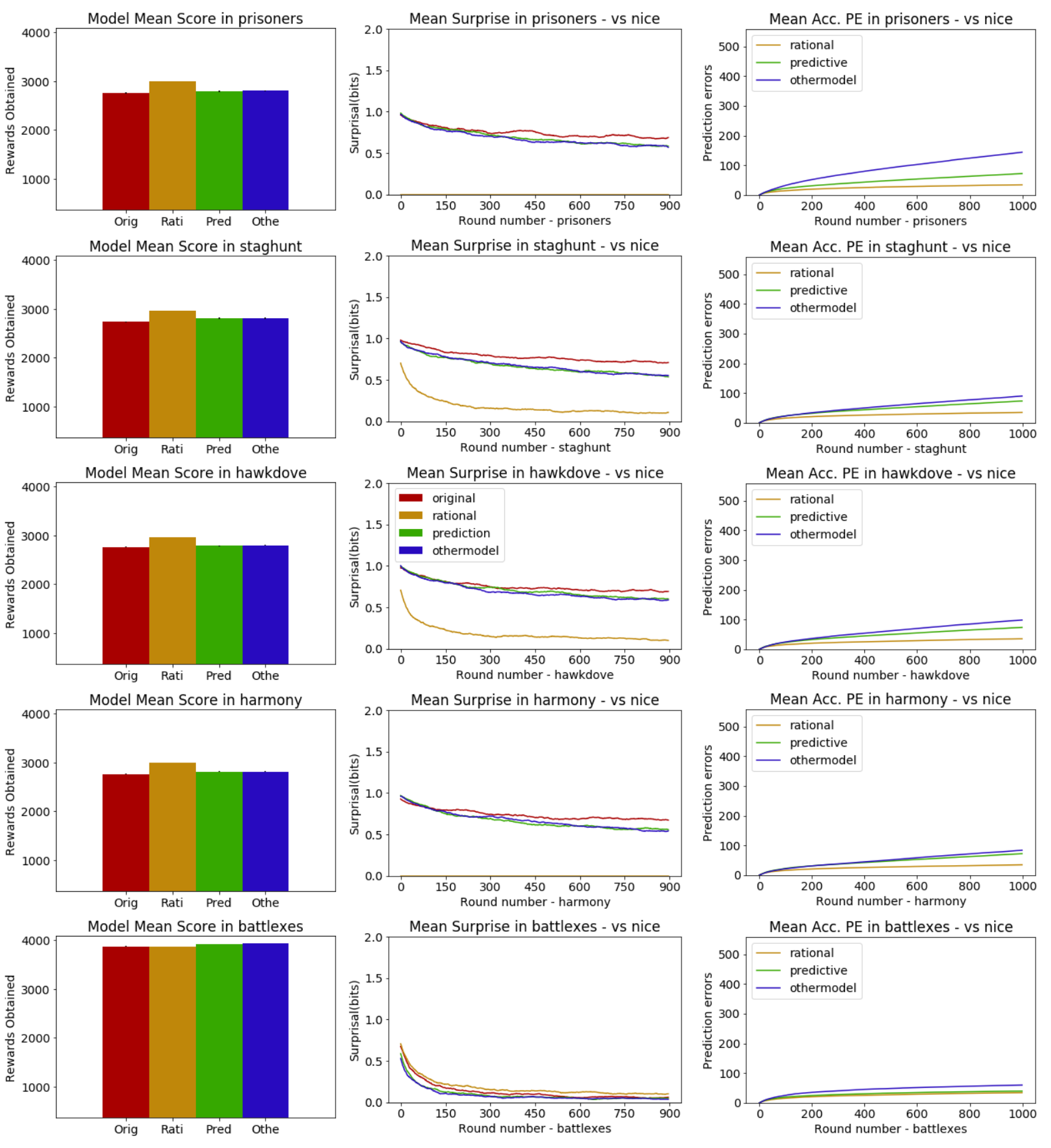}
\caption{Main results of the four behavioral models (Original, Rational, Predictive and Other's) against a Nice agent. The legend in the center applies to the left and middle columns.}
\label{results_nice}
\end{figure*}

In this experiment, the overall efficacy of all models has increased significantly compared to that of the previous experiment (see Figure \ref{results_nice}). The explanation was again that all models were able to successfully learn the optimal policy against a simple deterministic strategy such as the one exhibited by the Nice agent. But in this case, the best response strategy against a Nice agent is the one that renders the highest possible benefit in all games, as opposed to the best response strategy against the Greedy model, that in many occasions was sub-optimal. In terms of surprisal, we see again how the Predictive and the Other's-Model perform slightly better than the Original. This time the difference is smaller, but since predicting the opponent's behavior, in this case, is almost trivial, we don't find this result unexpected. Again we can also observe how the Original, the Predictive and the Other's-Model differ from optimal by comparing their convergence rate with that of the Rational agent. 

\subsection{Experiment 3: Against a Tit-for-tat agent}
\begin{figure*}[ht]
\centering
\includegraphics[width=.90\textwidth]{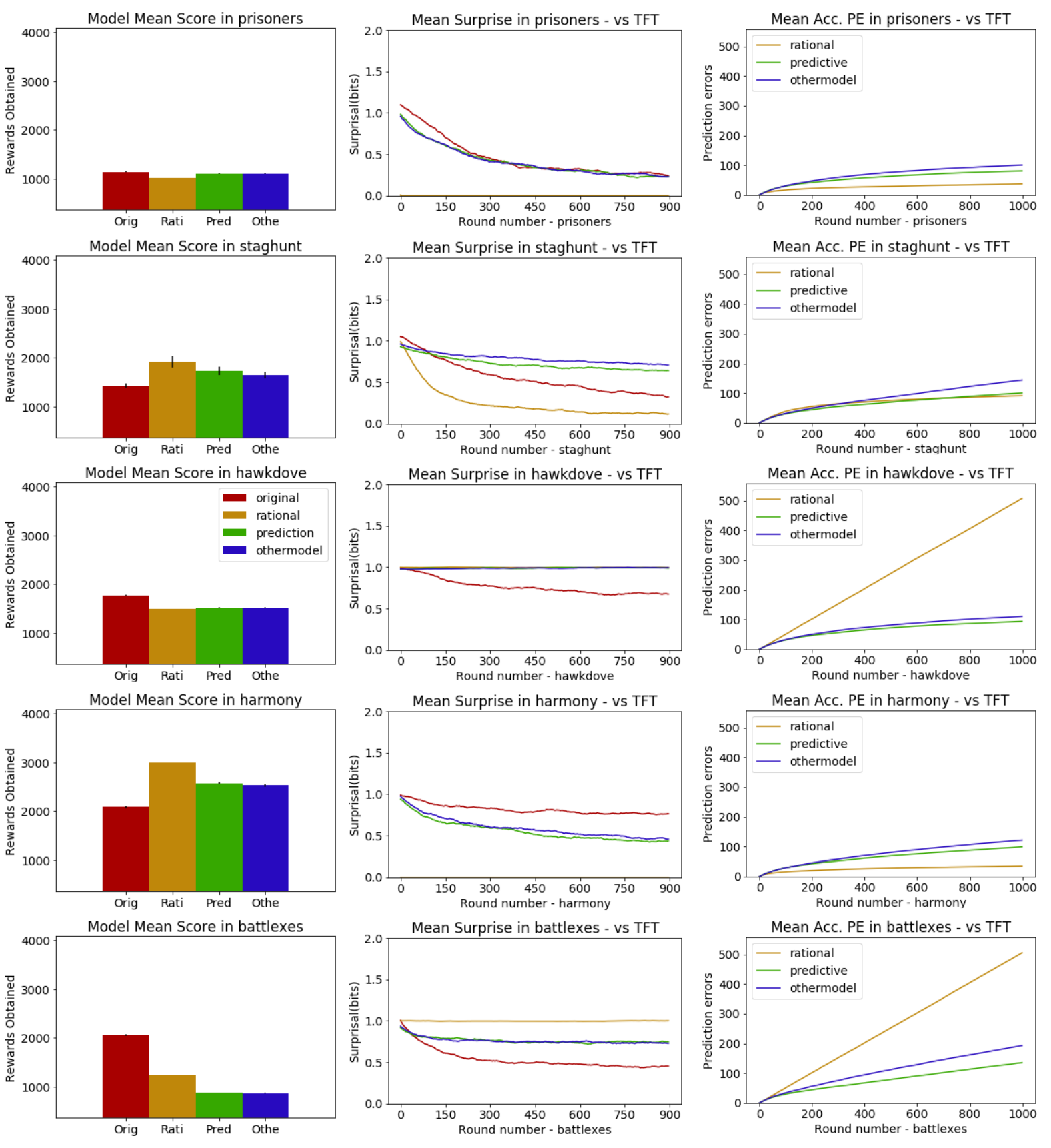}
\caption{Main results of the four behavioral models (Original, Rational, Predictive and Other's) against a TFT agent.}
\label{results_TFT}
\end{figure*}

The results of the models against the Tit-for-tat agent show a significant amount of variability in terms of efficacy (see Figure \ref{results_TFT}). The most remarkable result in this aspect is the superior efficacy of the non-predictive Original model in the Battle of the Exes. This model is able to beat the more complex predictive models in this case precisely because the predictive models are having problems to accurately predict the TFT agent. This accumulated error in prediction, in turn, drives their behavior toward a more unstable regime than the one achieved by the Original model. Moreover, given that this model has a more reduced state-space than the Predictive models and it only 'cares' about the previous state of the game, it capitalizes better than the rest on the anti-coordination structure of the Battle of the Exes.


In terms of suprisal, predictive models perform better than the Original in the Prisoners Dilemma and in the Harmony game, but this result is reversed in the Stag-Hunt and the Hawk-Dove and the Battle of the Exes, where the Original reaches lower levels of surprisal. We can see how in the cases that the game has one pure Nash equilibrium (the Prisoner's dilemma and the Harmony game), the predictive models perform as well in terms of surprisal as against the simpler deterministic agents studied before. However, in games with a wider variety of equilibria such as the Stag-hunt or the Hawk-dove, predictive models have overall more difficulties to fall into a stable state with the TFT opponent

Finally, the most salient feature regarding the resutls of Prediction accuracy are the elevated accumulation of errors of the Rational model in the Stag-Hunt and the Battle of the Exes, that almost reach 50\%. These results may seem counter-intuitive, but actually, they highlight one of the weak points of the predictive models and the strength of such a simple strategy. As stated in the model's description, the TFT agent selects its action based on the last action performed by its opponent, so attempting to predict a policy of such agent can become impossible for a predictive agent unless the dyad falls into an equilibrium state very soon or if the predictive agent was able to integrate it's own action into its prediction algorithm. 
\subsection{Experiment 4: Against the Original agent }
\begin{figure*}[ht]
\centering
\includegraphics[width=.90\textwidth]{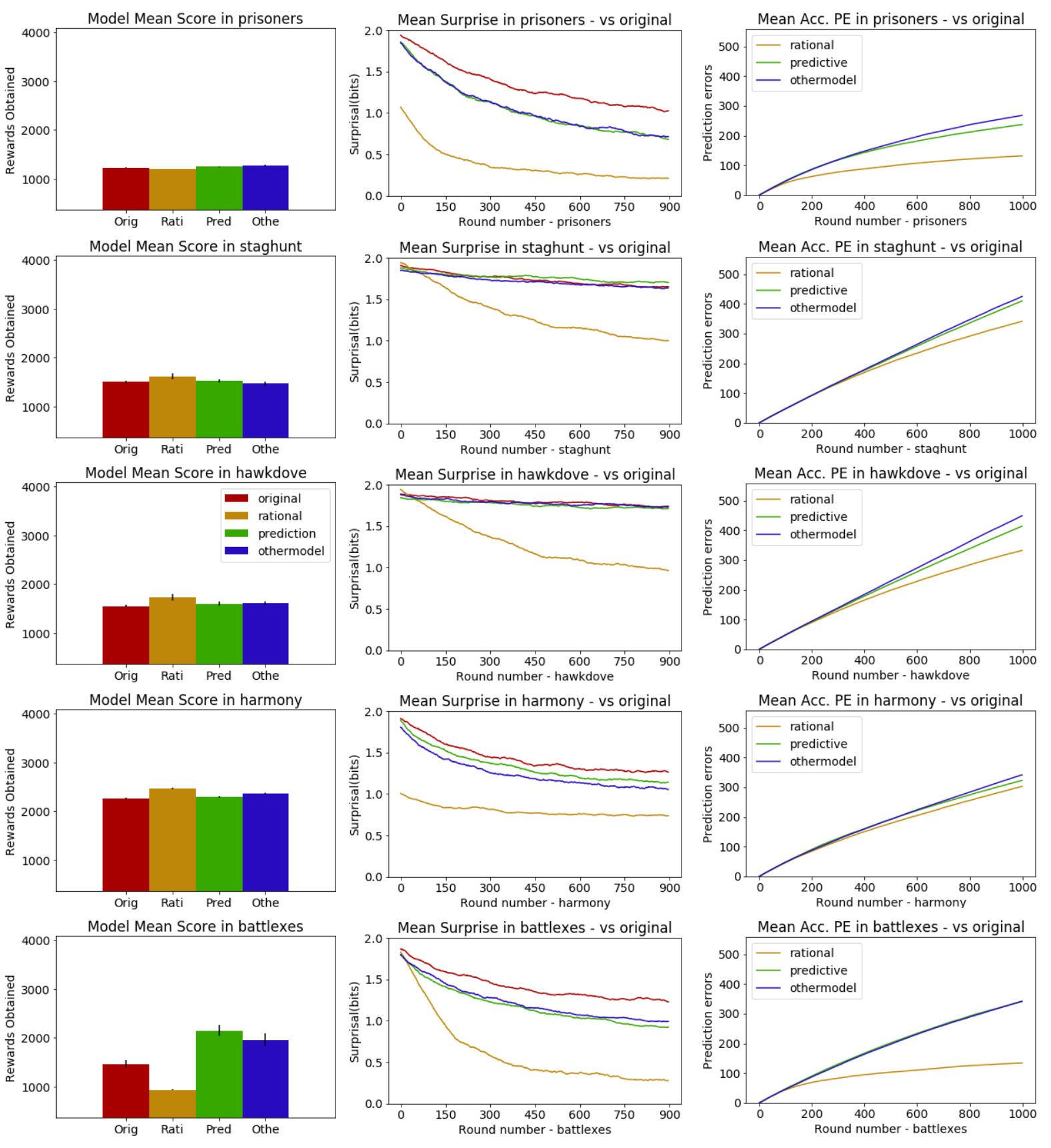}
\caption{Mean results of the four behavioral models (Original, Rational, Predictive and Other's) against the Original agent.}
\label{results_RL}
\end{figure*}

In this experiment, we also observe a similar performance of the models in terms of their efficacy to obtain rewards, with the exception of the Battle of the Exes, where the Predictive an Other's-Model achieve the best results (see Figure \ref{results_RL}. Remarkably, the Rational model is achieving a lower score than the rest. This is due to the fact that this model has fallen in a pure-dominance equilibrium with the Original model, where the best response for him is to perform the low-rewarding action. Again, in this case, the Original model capitalizes in the fact that is solely driven by its own self-utility maximization function, so it got the initiative to go to the higher-rewarding action faster than the Rational model. This, on the other hand, is predicting accurately and responding in the best optimal way to that accurate prediction. But in this case, this behavioral strategy makes it fall victim to the pure-dominance equilibrium enforced by its opponent. 

Regarding stability, the results show that overall the Predictive and the Other's-Model converge faster than the Original towards an equilibrium. In all cases, we can observe that these two predictive models fall in between the 'optimal' level of convergence of the Rational model and the more unstable level of the Original non-predictive model. The results of the Prediction accuracy show the worse overall performance of the four experiments studied so far. This, however, was an expected outcome since the predictive agents were facing a learning non-deterministic agent which is objectively more difficult to predict.

\subsection{Experiment 5: Continuous-Time Dyads }
\begin{figure*}[ht]
\centering
\includegraphics[width=.95\textwidth]{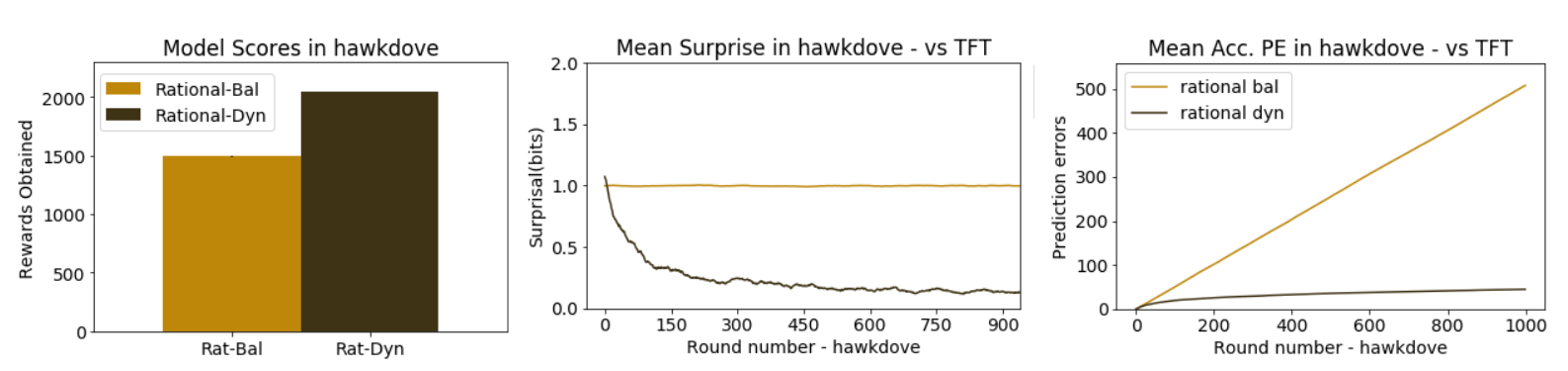}
\caption{Results of a game played in the continuous/real-time version of the game compared to the same game played in discrete-time.}
\label{results_realtime}
\end{figure*}

As an additional control experiment, here we test whether specific instances in experiment 3 above, where the Rational agent shows huge prediction errors against the TFT  agent, are purely a result of the discrete-time nature of ballistic games, or similar errors against the TFT agent also repeat under continuous-time interactions involving real-time feedback. As shown in Figure \ref{results_realtime}, this issue of prediction errors is in fact solved under real-time conditions. This is because the Rational agent is now able to change its chosen course of action in real-time before the round ends. This significantly reduces the prediction error of the rational agent against the TFT agent and enables the former to choose a better strategy, which consequently, drives the dyad to a stable equilibrium.

\section{Discussion}
This work demonstrates a novel control-based cognitive approach to autonomous multi-agent learning models, which combines adaptive feedback control with reinforcement learning. Based on a layered-control architecture, we have designed and implemented agent models characterized by seven different behavioral phenotypes. This includes both, deterministic agents as well as probabilistic learning agents. The former includes cooperative, greedy and tit-for-tat agents, whereas, the latter includes a reinforcement learning, rational, predictive and other's-model agents. We tested these agent models in dyadic games against each other across five different game theory settings (Harmony game, Stag-Hunt, Hawk-Dove, Prisoner's Dilemma and Battle of the Exes). From human behavioral experiments, it is known that the scope of these games sufficiently encompasses behavioral phenotypes in human social decision-making \cite{hawkins2016formation,poncela2016humans}. Our proposed agent models were designed and implemented with a view to simulating and understanding how cognitive agents learn and make decisions when interacting with other agents in social scenarios (in the context of the game-theoretic tasks stated above). 

Additionally, we also implemented our agent models as embodied ePuck robots. We tested these simulated robots within a spatial continuous version of the game involving real-time feedback. We showed that under fully embodied and situated conditions, limitations encountered by agents in ballistic scenarios (such as lack of convergence to an optimal solution) can overcome. Our simulated robots are Braitenberg Vehicles equipped with sensors and fully embodied controllers (which are the proposed control-based learning modules). These agents are situated in the sense that they only have partial observability of the environment and other agents depending on their location and sensors during rounds.


We found that pure reinforcement learning was inadequate in many tasks (as shown in performance metrics). On the other hand, agent models that predict actions or policies of opposing agents showed faster convergence to  equilibria and increased performance.  Interestingly, the rational agent (which knows what is the best action to take in each game and situation if the other's action is foreseen) outperforms others in many games. Our results show that agents using pure reinforcement-learning strategies are not optimal in most of the tested games. Additionally, upon examining human data obtained in the Battle of the Exes game by \cite{hawkins2016formation}, we found (not shown in the results) that surprisal profiles of human players matched closest to our predictive and other's-model agents, rather than the rational or pure RL agents.  

Of particular significance is the result that our predictive and other's-model agents correctly learn to predict actions and policies respectively of the opposing agent across all games. This shows how autonomous, embodied and situated multi-agent systems can be equipped with simple theory of mind capabilities (in the context of the specified tasks). For future work, it might be interesting to equip our agents with meta-learning models, so that they may be able to choose which behavioral phenotype to enact in a given social scenario and flexibly switch in changing environments.


Even though the different behavioral strategies modeled in this study consist some of the simplest building blocks of human behavior, it is by no means a complete list. Also, typical human behavior is not rigidly tied to a given model, but rather can flexibly shift between behavioral strategies. Model flexibility is definitely an important consideration for a ToM. Another issue that is relevant for many real-world tasks is the cost or penalty of being engaged too often in less desirable outcomes such as ties or low rewards. In a sense, this could serve as an incentive to improve performance in complex tasks and should be taken into account in a ToM. Yet another typical human feature is trust. Humans tend to form social conventions by agreeing on simple rules or deals and forming a trust relation. At the moment, our models currently lack this ability to form teams that agree on collective goals and work cooperatively to achieve success.

Finally, our control-based reinforcement learning architecture provides an interesting possibility to implement complex social behaviors on  multi-agent robotic platforms, such as humanoid robots  \cite{moulin2017embodied} or other socially intelligent artificial systems. Another scientific domain that may benefit from the type of agent models described in this paper, is evolutionary dynamics; in particular, the evolution of cognitive and intelligent agents. Social interactions involving cooperation and competition are known to play a key role in many evolutionary accounts of biological life and consciousness  \cite{arsiwalla2016consciousness},   
\cite{arsiwalla2017morphospace}.  Lastly, autonomous multi-agent models, of the type described above, with ToM capabilities, provide useful tools for modeling the dynamics of global socio-political and cultural phenomena.

\section*{Acknowledgments}
The research leading to these results has received funding from the European Commission’s Horizon 2020 socSMC project (socSMC-641321H2020) and by the European Research Council’s CDAC project   (ERC-2013-ADG341196).

\section*{Additional information}
The authors declare that they have no competing interests.

%
%
%
%






\section*{References}

\bibliography{main}

\begin{thebibliography}{10}
\expandafter\ifx\csname url\endcsname\relax
  \def\url#1{\texttt{#1}}\fi
\expandafter\ifx\csname urlprefix\endcsname\relax\def\urlprefix{URL }\fi
\expandafter\ifx\csname href\endcsname\relax
  \def\href#1#2{#2} \def\path#1{#1}\fi

\bibitem{premack1978does}
D.~Premack, G.~Woodruff, Does the chimpanzee have a theory of mind?, Behavioral
  and brain sciences 1~(4) (1978) 515--526.

\bibitem{baron1985does}
S.~Baron-Cohen, A.~M. Leslie, U.~Frith, Does the autistic child have a
  “theory of mind”?, Cognition 21~(1) (1985) 37--46.

\bibitem{premack1990infant}
D.~Premack, The infant's theory of self-propelled objects, Cognition 36~(1)
  (1990) 1--16.

\bibitem{lanctot2017unified}
M.~Lanctot, V.~Zambaldi, A.~Gruslys, A.~Lazaridou, K.~Tuyls, J.~P{\'e}rolat,
  D.~Silver, T.~Graepel, A unified game-theoretic approach to multiagent
  reinforcement learning, in: Advances in Neural Information Processing
  Systems, 2017, pp. 4190--4203.

\bibitem{lerer2018learning}
A.~Lerer, A.~Peysakhovich, Learning social conventions in markov games, arXiv
  preprint arXiv:1806.10071.

\bibitem{rabinowitz2018machine}
N.~C. Rabinowitz, F.~Perbet, H.~F. Song, C.~Zhang, S.~Eslami, M.~Botvinick,
  Machine theory of mind, arXiv preprint arXiv:1802.07740.

\bibitem{schmidhuber2015deep}
J.~Schmidhuber, Deep learning in neural networks: An overview, Neural networks
  61 (2015) 85--117.

\bibitem{yoshida2008game}
W.~Yoshida, R.~J. Dolan, K.~J. Friston, Game theory of mind, PLoS computational
  biology 4~(12) (2008) e1000254.

\bibitem{baker2011bayesian}
C.~Baker, R.~Saxe, J.~Tenenbaum, Bayesian theory of mind: Modeling joint
  belief-desire attribution, in: Proceedings of the annual meeting of the
  cognitive science society, Vol.~33, 2011.

\bibitem{baker2017rational}
C.~L. Baker, J.~Jara-Ettinger, R.~Saxe, J.~B. Tenenbaum, Rational quantitative
  attribution of beliefs, desires and percepts in human mentalizing, Nature
  Human Behaviour 1~(4) (2017) 0064.

\bibitem{tenenbaum2018building}
J.~Tenenbaum, Building machines that learn and think like people, in:
  Proceedings of the 17th International Conference on Autonomous Agents and
  MultiAgent Systems, International Foundation for Autonomous Agents and
  Multiagent Systems, 2018, pp. 5--5.

\bibitem{albrecht2018autonomous}
S.~V. Albrecht, P.~Stone, Autonomous agents modelling other agents: A
  comprehensive survey and open problems, Artificial Intelligence 258 (2018)
  66--95.

\bibitem{freire2018modeling}
I.~T. Freire, C.~Moulin-Frier, M.~Sanchez-Fibla, X.~D. Arsiwalla, P.~Verschure,
  Modeling the formation of social conventions in multi-agent populations,
  arXiv preprint arXiv:1802.06108.

\bibitem{freire2018limits}
I.~T. Freire, J.-Y. Puigb{\`o}, X.~D. Arsiwalla, P.~F. Verschure, Limits of
  multi-agent predictive models in the formation of social conventions,
  Artificial Intelligence Research and Development: Current Challenges, New
  Trends and Applications 308 (2018) 297.

\bibitem{kleiman2016coordinate}
M.~Kleiman-Weiner, M.~K. Ho, J.~L. Austerweil, M.~L. Littman, J.~B. Tenenbaum,
  Coordinate to cooperate or compete: abstract goals and joint intentions in
  social interaction, in: CogSci, 2016.

\bibitem{perolat2017multi}
J.~Perolat, J.~Z. Leibo, V.~Zambaldi, C.~Beattie, K.~Tuyls, T.~Graepel, A
  multi-agent reinforcement learning model of common-pool resource
  appropriation, in: Advances in Neural Information Processing Systems, 2017,
  pp. 3643--3652.

\bibitem{peysakhovich2018prosocial}
A.~Peysakhovich, A.~Lerer, Prosocial learning agents solve generalized stag
  hunts better than selfish ones, in: Proceedings of the 17th International
  Conference on Autonomous Agents and MultiAgent Systems, International
  Foundation for Autonomous Agents and Multiagent Systems, 2018, pp.
  2043--2044.

\bibitem{freire2018modelingb}
I.~T. Freire, J.-Y. Puigb{\`o}, X.~D. Arsiwalla, P.~F. Verschure, Modeling the
  opponent’s action using control-based reinforcement learning, in:
  Conference on Biomimetic and Biohybrid Systems, Springer, 2018, pp. 179--186.

\bibitem{gasparrini2018loss}
M.~J. Gaparrini, M.~S{\'a}nchez-Fibla, Loss aversion fosters coordination in
  independent reinforcement learners, Artificial Intelligence Research and
  Development: Current Challenges, New Trends and Applications 308 (2018) 307.

\bibitem{leibo2017multi}
J.~Z. Leibo, V.~Zambaldi, M.~Lanctot, J.~Marecki, T.~Graepel, Multi-agent
  reinforcement learning in sequential social dilemmas, in: Proceedings of the
  16th Conference on Autonomous Agents and MultiAgent Systems, International
  Foundation for Autonomous Agents and Multiagent Systems, 2017, pp. 464--473.

\bibitem{peysakhovich2017consequentialist}
A.~Peysakhovich, A.~Lerer, Consequentialist conditional cooperation in social
  dilemmas with imperfect information, arXiv preprint arXiv:1710.06975.

\bibitem{nash1950equilibrium}
J.~F. Nash, et~al., Equilibrium points in n-person games, Proceedings of the
  national academy of sciences 36~(1) (1950) 48--49.

\bibitem{hawkins2016formation}
R.~X. Hawkins, R.~L. Goldstone, The formation of social conventions in
  real-time environments, PloS one 11~(3) (2016) e0151670.

\bibitem{hawkins2018emergence}
R.~X. Hawkins, N.~D. Goodman, R.~L. Goldstone, The emergence of social norms
  and conventions, Trends in cognitive sciences.

\bibitem{poncela2016humans}
J.~Poncela-Casasnovas, M.~Guti{\'e}rrez-Roig, C.~Gracia-L{\'a}zaro, J.~Vicens,
  J.~G{\'o}mez-Garde{\~n}es, J.~Perell{\'o}, Y.~Moreno, J.~Duch,
  A.~S{\'a}nchez, Humans display a reduced set of consistent behavioral
  phenotypes in dyadic games, Science advances 2~(8) (2016) e1600451.

\bibitem{sanfey2007social}
A.~G. Sanfey, Social decision-making: insights from game theory and
  neuroscience, Science 318~(5850) (2007) 598--602.

\bibitem{verschure2003environmentally}
P.~F. Verschure, T.~Voegtlin, R.~J. Douglas, Environmentally mediated synergy
  between perception and behaviour in mobile robots, Nature 425~(6958) (2003)
  620.

\bibitem{moulin2016top}
C.~Moulin-Frier, X.~D. Arsiwalla, J.-Y. Puigb{\`o}, M.~Sanchez-Fibla, A.~Duff,
  P.~F. Verschure, Top-down and bottom-up interactions between low-level
  reactive control and symbolic rule learning in embodied agents., in:
  CoCo@NIPS, 2016.

\bibitem{braitenberg1986vehicles}
V.~Braitenberg, Vehicles: Experiments in synthetic psychology, MIT press, 1986.

\bibitem{corbetta2002control}
M.~Corbetta, G.~L. Shulman, Control of goal-directed and stimulus-driven
  attention in the brain, Nature reviews neuroscience 3~(3) (2002) 201.

\bibitem{koechlin2003architecture}
E.~Koechlin, C.~Ody, F.~Kouneiher, The architecture of cognitive control in the
  human prefrontal cortex, Science 302~(5648) (2003) 1181--1185.

\bibitem{munakata2011unified}
Y.~Munakata, S.~A. Herd, C.~H. Chatham, B.~E. Depue, M.~T. Banich, R.~C.
  O’Reilly, A unified framework for inhibitory control, Trends in cognitive
  sciences 15~(10) (2011) 453--459.

\bibitem{den2012prediction}
H.~E. Den~Ouden, P.~Kok, F.~P. De~Lange, How prediction errors shape
  perception, attention, and motivation, Frontiers in psychology 3 (2012) 548.

\bibitem{wacongne2011evidence}
C.~Wacongne, E.~Labyt, V.~van Wassenhove, T.~Bekinschtein, L.~Naccache,
  S.~Dehaene, Evidence for a hierarchy of predictions and prediction errors in
  human cortex, Proceedings of the National Academy of Sciences 108~(51) (2011)
  20754--20759.

\bibitem{sutton1988learning}
R.~S. Sutton, Learning to predict by the methods of temporal differences,
  Machine learning 3~(1) (1988) 9--44.

\bibitem{axelrod1981evolution}
R.~Axelrod, W.~D. Hamilton, The evolution of cooperation, science 211~(4489)
  (1981) 1390--1396.

\bibitem{axelrod1980effective}
R.~Axelrod, Effective choice in the prisoner's dilemma, Journal of conflict
  resolution 24~(1) (1980) 3--25.

\bibitem{shannon1948mathematical}
C.~E. Shannon, A mathematical theory of communication, Bell system technical
  journal 27~(3) (1948) 379--423.

\bibitem{moulin2017embodied}
C.~Moulin-Frier, J.-Y. Puigbo, X.~D. Arsiwalla, M.~Sanchez-Fibla, P.~Verschure,
  Embodied artificial intelligence through distributed adaptive control: An
  integrated framework, ArXiv e-prints.

\bibitem{arsiwalla2016consciousness}
X.~D. Arsiwalla, I.~Herreros, C.~Moulin-Frier, M.~S{\'a}nchez-Fibla, P.~F.
  Verschure, Is consciousness a control process?, in: CCIA, 2016, pp. 233--238.

\bibitem{arsiwalla2017morphospace}
X.~D. Arsiwalla, R.~Sole, C.~Moulin-Frier, I.~Herreros, M.~Sanchez-Fibla,
  P.~Verschure, The morphospace of consciousness, arXiv preprint
  arXiv:1705.11190.

\end{thebibliography}

\end{document}